%% file: main_manuscript.tex
\documentclass[sigconf,screen,nonacm]{acmart}
\AtBeginDocument{%
  }

\setcopyright{acmlicensed}
\copyrightyear{2026}
\acmYear{2026}
\acmDOI{XXXXXXX.XXXXXXX}

\acmConference[ACM MobiSys '2026]{ACM MobiSys}{June 21--25,
  2026}{Cambridge, UK}
\acmBooktitle{Proceedings of the 24th International Conference on Mobile Systems, Applications, and Services (MobiSys '26), June 21--25, 2026, Cambridge, UK}

\usepackage{subfig}

\begin{document}

\title{ML and Smartphones Assisted Real-Time Uplink Performance Prediction in 5G Cellular System}

 \author{Md Mahfuzur Rahman}
  \affiliation{%
  \institution{Virginia Polytechnic Institute and State University}
   \city{Blacksburg}
   \state{Virginia}
   \country{USA}
 }
 \email{mrahman2@vt.edu}
 
 \author{Jareen Shuva}
   \affiliation{%
  \institution{Virginia Polytechnic Institute and State University}
   \city{Blacksburg}
   \state{Virginia}
   \country{USA}
 }
 \email{jareenshuva@vt.edu}
 
 \author{Nishith Tripathi}
   \affiliation{%
  \institution{Virginia Polytechnic Institute and State University}
   \city{Blacksburg}
   \state{Virginia}
   \country{USA}
 }
 \email{nishith@vt.edu}
 \author{Lingjia Liu}
    \affiliation{%
  \institution{Virginia Polytechnic Institute and State University}
   \city{Blacksburg}
   \state{Virginia}
   \country{USA}
 }
 \email{ljliu@vt.edu}
 \author{Jeffrey Reed}
\affiliation{%
  \institution{Virginia Polytechnic Institute and State University}
   \city{Blacksburg}
   \state{Virginia}
   \country{USA}
 }
\email{reedjh@vt.edu}

\renewcommand{\shortauthors}{Md Mahfuzur Rahman et al.}

\begin{abstract}
We propose a machine learning (ML) and smartphone-assisted framework for uplink performance prediction in a private, realistic 5G cellular system using real-time measurements in both indoor and outdoor settings. This work presents a comprehensive data-driven evaluation of 5G performance prediction using a controllable software-defined radio test environment. The experimental platform is built on srsRAN’s 5G NR stack running on a Dell workstation configured as a gNB and 5G core operating at 3.4 GHz. Two commercial Google Pixel 7a devices are instrumented to capture uplink metrics, including channel quality indicator (CQI), modulation and coding scheme (MCS), throughput, transmission time interval (TTI), and block error rate (BLER). Different types of traffic are generated using industry-standard tools such as Ookla and iperf, spanning stationary, pedestrian, and mobility cases under both line-of-sight (LOS) and non-line-of-sight (nLOS) propagation environments. Additional datasets include YouTube video sessions and global server endpoints to introduce variability in path characteristics. The resulting measurements, including multi-UE interference conditions, serve as training data for several supervised regression models. Five learning algorithms—linear regression, decision tree, random forest, XGBoost, and LightGBM—are benchmarked for prediction accuracy. The study shows that reliable forecasting of throughput and BLER is feasible using only COTS smartphones and widely available ML methods, offering a practical pathway for real-world 5G network performance estimation.

\end{abstract}





\maketitle




\section{Background and Related Research}
\subsection{Introduction}

\input{sections/intro}
\subsection{Related Research}

\input{sections/related_work}

\section{Motivation and Contribution}

\input{sections/motivation}

\input{sections/contrib}

\section{Machine Learning} 
\input{sections/ml1}


\section{Regression Model Formulations}

\input{sections/ml}

\section{System Model and Traffic Generation}

\input{sections/system}

\section{Result and Analysis} 

\input{sections/simulation}


\section{Conclusion}
\input{sections/conclusion}

\bibliographystyle{ACM-Reference-Format}
\bibliography{oran}










\end{document}

%% file: sections/intro.tex
The emergence of the Open Radio Access Network (O-RAN) paradigm has significantly transformed the modern cellular landscape by promoting openness, interoperability, software-driven design, and AI-enabled control mechanisms \cite{tripathi2025fundamentals}. Although fixed networks have long benefited from the principles of Software-Defined Networking (SDN), fully realizing these innovations within mobile systems remains an ongoing challenge. Achieving a robust and intelligent O-RAN deployment requires addressing several technical hurdles, including the acquisition of high-quality datasets for ML algorithms, building experimental environments that are both repeatable and minimally intrusive, automating continuous software validation, and maintaining seamless coordination across diverse network components \cite{Colosseum}.

Experimental open-source software stacks—most notably OpenAirInterface (OAI) and the Software Radio Systems RAN (srsRAN)—have become central tools for 5G New Radio (NR) research and prototyping. These platforms implement the full 3GPP protocol stack and operate with Commercial-Off-The-Shelf (COTS) UEs and SDR hardware, enabling cost-effective and flexible experimentation \cite{openairinterface, srsRAN4G}. Their modular architectures allow researchers to rapidly test new features, adjust network configurations, and conduct controlled measurements. In particular, srsRAN has gained substantial traction due to its reproducibility, lightweight deployment, and ease of integration with machine-learning pipelines.

As 5G systems continue to target extreme reliability and very low latency, precise estimation of network performance metrics has become critical for intelligent resource management. Throughput in cellular systems is tightly coupled with various PHY-layer characteristics, including Modulation and Coding Scheme (MCS), Signal-to-Interference-plus-Noise Ratio (SINR), and Reference Signal Received Power (RSRP) \cite{perveen2022dynamic, teixeira2021predictive}. Machine-learning techniques have therefore been increasingly adopted to capture these complex nonlinear relationships and provide real-time bit rate forecasts from instantaneous radio measurements \cite{eyceyurt2022machine, rehmani2023machine, elsherbiny2020throughput}. Such predictive tools support enhanced network automation by informing scheduling, mobility management, and link adaptation based on richer channel state information (CSI).

srsRAN testbeds have played a major role in the advancement of ML-based throughput prediction. Their fine-grained control of network parameters and ability to reproduce experiments make them ideal for assessing bit rate prediction models under realistic yet repeatable configurations \cite{alves2024experimental, Ye}. Prior work has shown that algorithms such as k-Nearest Neighbor and Random Forest Regression can achieve $R^2$ values beyond 0.9 in srsRAN-based LTE setups when predicting uplink or downlink bit rate \cite{elsherbiny2020throughput, eyceyurt2022machine}. Beyond throughput estimation, the same platform has supported studies in areas such as wireless security \cite{Shen}, protocol fuzzing \cite{mishra2023scaling}, and diagnostic analytics \cite{forbes2023closer}.

Several influential contributions illustrate the breadth of research in this area. Koenig et al. \cite{koenig2021throughput} introduced a cross-technology prediction model applicable to both LTE and 5G. Rehmani et al. \cite{rehmani2023machine} validated Random Forest and regression models with real operator data, demonstrating ML’s strong potential for adaptive network management. Dias et al. \cite{dias2022rsrp2} used an srsRAN platform to estimate RSRP values for LTE planning and coverage analysis. Sinha et al. \cite{sinha2022prototyping} presented a flexible SDR framework combining OAI and srsRAN to support multi-generation experimentation within emerging O-RAN architectures.

Multiple ML models have been applied to 5G bit rate forecasting, including LightGBM, XGBoost, Random Forest, Decision Tree, and Linear Regression. Linear Regression has served as a baseline in many studies for capturing simple relations between fundamental radio parameters. Elsherbiny et al. \cite{elsherbiny20204g} illustrated that, despite its simplicity, linear modeling can yield real-time throughput estimates from COTS UE measurements. Decision Tree Regression, known for its interpretability, has been employed to model nonlinear dependencies and predict inter-frequency indicators vital to mobility optimization \cite{sonnert2018predicting}. Ensemble approaches such as Random Forest often outperform single-model predictors; for example, Chang and Baliga \cite{chang2021development} demonstrated its effectiveness for path-loss estimation in dense urban settings, and smartphone-derived RSRP prediction has also benefited from its robustness.

Using COTS smartphones for data collection underscores the real-world relevance of predictive modeling in operational 5G environments. These devices simplify large-scale measurement campaigns and, when paired with open-source platforms like srsRAN, enable highly flexible research testbeds bridging experimental and commercial deployment contexts. In this work, we build on these advancements by developing and validating an srsRAN-based real-time LTE measurement platform using commercial smartphones. Rather than performing direct PHY-layer optimization, our goal is to estimate downlink bit rate and SNR using passively observed measurements. We develop an efficient ML pipeline combining Linear Regression, Decision Tree Regression, Random Forest Regression, XGBoost, and LightGBM, trained on live network features including transmission time interval (TTI), bit rate, MCS, SNR, and BLER. This approach enables non-intrusive estimation of bit rate and SNR, supporting proactive QoS optimization, scheduling, and link adaptation in emerging wireless systems. It is important to emphasize that these ML-driven estimators do not modify underlying physical-layer capabilities; rather, they provide an intelligent layer for early performance inference.

%% file: sections/related_work.tex
Recent efforts in cellular-performance modeling have applied machine learning to bit rate prediction; however, most prior studies rely on measurements taken from operational commercial networks, where researchers have limited control over experimental parameters and network behavior.

A substantial body of work has focused on drive-test–based data collection. Al-Thaedan et al. \cite{Al-Thaedan} examined downlink throughput prediction in urban LTE environments using datasets from three operators. Their evaluation of K-Nearest Neighbors, Decision Trees, Support Vector Regression, and Linear Regression—using features such as GPS location, SINR, RSRQ, and RSRP—reported R² values approaching 99

Measurements gathered directly from commercial networks have also been used to build predictive models. Minovski et al. \cite{Minovski} developed throughput predictors for both LTE and 5G based on SINR, RSRQ, and RSRP across rural, urban, and underground environments. Utilizing custom TWAMP probes and TEMS Pocket tools for data collection, they demonstrated that reliable bit rate labeling can be achieved without controlled testbed hardware. Basit et al. \cite{Basit} extended this methodology through a large-scale campaign—spanning 1000 km—over commercial LTE and 5G networks using Android devices and Accuver tools. Their comparison of Gradient Boosting and Multi-Layer Perceptron models showed that accuracy deteriorated at 100 ms granularity for application-level metrics, largely due to rapidly fluctuating commercial-network dynamics and inconsistencies in measurement sequences.

Deep learning has further been explored for modeling highly variable throughput patterns. Raca et al. \cite{Raca} built a hybrid ML/DL prediction engine leveraging both network-side and device-side metrics for LTE. Their results demonstrated that LSTM networks outperform SVM and Random Forest models under highly dynamic conditions. They also validated the practical utility of prediction-assisted adaptation, showing enhanced video streaming QoE via Android APIs and custom instrumentation. A technical report from Clemson University \cite{clemson2022machine} corroborated these findings, showing that deeper neural architectures more effectively capture nonlinear throughput behavior compared to classical ML methods. Abdiel \cite{abdiel2022forecasting} proposed an LTE forecasting framework aimed at both operational optimization and strategic business analytics.

Despite these notable contributions, most published techniques depend on live operator infrastructure or drive-test campaigns, which inherently limit reproducibility and restrict the degree of control over radio parameters. In contrast, our work employs a fully customizable 5G experimentation platform built using srsRAN and COTS smartphones. This setup supports controlled, repeatable experiments, fine-grained tuning of radio parameters, and seamless integration with machine learning pipelines. We intentionally focus on lightweight, interpretable models—LGBM, XGBoost, Random Forest, Decision Trees, and Linear Regression—to achieve a desirable balance between accuracy and computational efficiency. Unlike prior deep-learning–based solutions, our framework enables real-time inference suitable for on-device and resource-constrained deployments. By implementing and validating these algorithms in a live programmable testbed, we provide a reproducible foundation for ML-driven SNR and bit rate prediction under tunable 5G network conditions.

%% file: sections/motivation.tex
The evolution of mobile networks toward greater openness, programmability, and embedded intelligence has elevated the Open Radio Access Network (O-RAN) framework as a pivotal architectural shift. By encouraging modular, software-driven, and interoperable designs, Open RAN lowers entry barriers and accelerates innovation across the wireless ecosystem. However, realizing this vision in production environments requires overcoming several foundational challenges, including ensuring high-quality datasets for model training, constructing non-intrusive experimental methodologies, and providing realistic environments for validating AI/ML algorithms.

Accurate prediction of channel metrics—particularly SNR and throughput—is critical for enabling intelligent scheduling, dynamic link adaptation, and maintaining QoS in 5G networks. Key physical-layer indicators such as MCS, SNR, BLER, and TTI directly affect achievable bit rate, yet the nonlinear and interdependent nature of these parameters makes them difficult to model using traditional heuristic approaches. SNR itself is shaped by a similar set of interacting factors, including MCS, BLER, TTI, and the resulting bit rate, further complicating analytical modeling.

Machine learning offers an effective means of capturing these hidden relationships by learning directly from measurement data. To rigorously evaluate such models, researchers require platforms that offer an appropriate blend of realism, flexibility, and controlled experimentation. Commercial operator networks do not satisfy these needs, as they restrict parameter visibility and limit experimental configurability. This motivates the need for low-cost, reproducible 5G research infrastructures that can support ML-driven network studies.

Open-source implementations like srsRAN, combined with Software Defined Radios (SDRs) and commercial off-the-shelf smartphones, provide a practical foundation for such investigations. These platforms enable end-to-end experimentation with fine-grained control over radio settings and network parameters. Although their use is expanding within the research community, the systematic integration of COTS UEs with srsRAN for real-time, ML-based throughput and SNR prediction using live data has not been thoroughly explored.

The present work fills this gap by illustrating how accessible hardware and open-source 5G standalone implementations can be leveraged to conduct reproducible, data-centric studies of network performance. Our methodology aligns closely with the Open RAN philosophy while addressing the practical needs of emerging 5G systems that increasingly rely on data-driven intelligence.

%% file: sections/contrib.tex
This study advances data-driven network intelligence by leveraging an open, fully configurable 5G platform. The key contributions of our work are summarized as follows:
\begin{itemize}
\item \textbf{Machine Learning Pipeline for Uplink SNR Prediction:} We implement and assess five regression techniques—Linear Regression, Random Forest, Decision Tree, LightGBM, and XGBoost—to estimate uplink SNR directly from UE-reported physical-layer indicators.

\item \textbf{Learning-Based Uplink Throughput Estimation:} Using the same set of models, we construct a prediction framework for uplink bit rate based on UE-side measurements. This enables passive inference of throughput without generating active traffic or modifying network behavior.

\item \textbf{Interpretable Feature Contribution Analysis:} Through detailed importance evaluation, we show that MCS consistently emerges as the most influential feature for both SNR and bit rate prediction. This provides transparent insights into the physical-layer relationships driving model performance.

\item \textbf{Building Blocks for Reinforcement Learning Control Mechanisms:} By jointly modeling channel quality (SNR) and achievable performance (bit rate), the proposed system forms a principled basis for defining RL state vectors. This design will serve as the initial stage for future RL-based control frameworks explored in our upcoming journal article.
\end{itemize}

%% file: sections/ml1.tex
Machine learning, a major subfield of artificial intelligence, enables computational systems to automatically uncover structure in data and make informed decisions without explicit rule-based programming. Broadly, ML techniques are grouped into three categories: supervised learning, which leverages labeled datasets to learn predictive functions; unsupervised learning, which uncovers latent patterns in unlabeled data; and reinforcement learning, where agents iteratively refine their strategies by interacting with an environment and receiving evaluative feedback. Collectively, these paradigms have transformed a diverse range of domains including wireless networking, computer vision, natural language understanding, and autonomous control.

\subsection{Supervised Learning}

Supervised learning constitutes one of the most widely used ML frameworks, where the goal is to learn a predictive model from labeled training examples. By observing input--output mappings, the model estimates a function capable of generalizing to new, unseen inputs. Numerous mobile and wireless applications employ supervised learning. For example, Zhao et al.~\cite{yao2017deepsense} introduced DeepSense, a unified deep-learning architecture for processing time-series data from mobile sensors, demonstrating strong performance in recognition and identification tasks using CNN and RNN layers.

\subsubsection{Mathematical Framework}

Let the training dataset be 
\[
\mathcal{D} = \{(x_1, y_1), \ldots, (x_N, y_N)\},
\]
where $x_i \in \mathcal{X}$ denotes a feature vector and $y_i \in \mathcal{Y}$ denotes the corresponding label. The learning problem aims to find a hypothesis $h : \mathcal{X} \rightarrow \mathcal{Y}$ that minimizes the expected (true) risk:

\begin{equation}
R(h) = \mathbb{E}_{(x,y) \sim P(X,Y)} [\ell(h(x), y)],
\end{equation}

where $\ell(\cdot)$ is a loss function and $P(X,Y)$ is the unknown data distribution. Since $P$ is unavailable, empirical risk minimization is performed:

\begin{equation}
\hat{R}(h) = \frac{1}{N}\sum_{i=1}^{N} \ell(h(x_i), y_i).
\end{equation}

For parametric models $f(\cdot; \theta)$, training involves solving:

\begin{equation}
\theta^* = \arg\min_{\theta} \left[
\frac{1}{N}\sum_{i=1}^{N} \ell(f(x_i; \theta), y_i)
+ \lambda R(\theta)
\right],
\end{equation}

where $R(\theta)$ is a regularization penalty and $\lambda$ controls its contribution.

\subsection{Reinforcement Learning}

Reinforcement learning (RL) describes a decision-making paradigm in which an agent interacts with an environment and refines its behavior to maximize cumulative reward over time~\cite{ramin}. Unlike supervised learning, RL does not require labeled datasets; instead, the agent learns from the consequences of its actions.

\subsubsection{Markov Decision Process Representation}

RL is commonly formulated as a Markov Decision Process (MDP), defined by the tuple $(\mathcal{S}, \mathcal{A}, P, R, \gamma)$, where:
\begin{itemize}
    \item $\mathcal{S}$: set of states,
    \item $\mathcal{A}$: set of actions,
    \item $P(s'|s,a)$: transition probability,
    \item $R(s,a,s')$: reward function,
    \item $\gamma \in [0,1)$: discount factor.
\end{itemize}

The agent seeks a policy $\pi : \mathcal{S} \rightarrow \mathcal{A}$ that maximizes the expected discounted return:

\begin{equation}
J(\pi) = \mathbb{E}_{\pi}\left[ \sum_{t=0}^{\infty} \gamma^t R(s_t, a_t, s_{t+1}) \right].
\end{equation}

\subsubsection{Value Functions}

The state-value function under policy $\pi$ is

\begin{equation}
V^\pi(s) = \mathbb{E}_{\pi}\left[\sum_{t=0}^{\infty}\gamma^t r_t \mid s_0 = s \right],
\end{equation}

while the action-value (Q-value) function is

\begin{equation}
Q^\pi(s,a) = \mathbb{E}_{\pi}\left[\sum_{t=0}^{\infty}\gamma^t r_t \mid s_0 = s, a_0 = a \right].
\end{equation}

The optimal variants satisfy the Bellman optimality relations:

\begin{equation}
V^*(s) = \max_a \left[ R(s,a) + \gamma \sum_{s'} P(s'|s,a)V^*(s') \right],
\end{equation}

\begin{equation}
Q^*(s,a) = R(s,a) + \gamma \sum_{s'} P(s'|s,a)\max_{a'}Q^*(s',a').
\end{equation}

\subsection{From Supervised Learning to Reinforcement Learning}

In modern wireless systems, numerous radio parameters—including scheduling indicators, power control variables, slicing configurations, and modulation characteristics—collectively influence key objectives such as throughput, latency, reliability, and energy efficiency. Determining which parameters most strongly affect these objectives is essential for designing effective RL-based network controllers. These influential parameters ultimately form the state representation supplied to the RL agent.

However, not all candidate features contribute meaningfully. Incorporating irrelevant or redundant variables may lead to unnecessary computational burden, slower convergence, and degraded learning performance. Consequently, performing a systematic feature importance assessment is vital for constructing a compact and informative RL state vector.

In this work, we carry out a supervised-learning–based feature importance analysis to evaluate the relative relevance of candidate radio parameters. The analysis examines how different features correlate with performance outcomes and how strongly they influence predictive models across diverse scenarios. The resulting rankings guide the selection of state variables for RL-based control in next-generation wireless systems.

This study represents the initial stage of a broader research framework. The feature set identified here will form the basis of the RL state representation in our subsequent journal publication, where we will conduct extensive simulation and real-time testbed experiments to design and validate complete RL-driven network optimization mechanisms for future 5G and beyond systems.

%% file: sections/ml.tex
This section outlines the theoretical principles and characteristics of the five regression models used to predict CQI and downlink bit rate in 5G networks: LightGBM, XGBoost, Random Forest, Decision Tree, and Linear Regression. The selection of these algorithms balances interpretability, computational efficiency, and predictive capacity across varying levels of model complexity.

\subsection{Random Forest Regression}

Random Forest is an ensemble technique that constructs multiple decision trees, each trained on bootstrapped data samples and randomly selected feature subsets. The final prediction is obtained by averaging the outputs of all trees:

\begin{equation}
\hat{y} = \frac{1}{B} \sum_{b=1}^{B} T_b(\mathbf{x}),
\end{equation}

where $T_b(\mathbf{x})$ denotes the $b$-th tree's prediction and $B$ is the ensemble size. This averaging reduces the variance typical of single decision trees and improves generalization.

Random Forest models are well suited for capturing nonlinear interactions among physical-layer variables such as bandwidth allocation, modulation configurations, and adaptive scheduling. They additionally provide feature importance scores that help identify dominant contributors to 5G performance.

\subsection{Decision Tree Regression}

Decision Tree Regression partitions the feature space into distinct regions through recursive binary splits. At each split, the algorithm selects a feature $j$ and threshold $t$ that minimize the variance of the target variable within the resulting partitions:

\begin{equation}
\min_{j, t} \left(
\sum_{x_i \in R_1(j,t)} (y_i - \bar{y}_{R_1})^2 +
\sum_{x_i \in R_2(j,t)} (y_i - \bar{y}_{R_2})^2
\right),
\end{equation}

where $R_1$ and $R_2$ are defined by $x_j \leq t$ and $x_j > t$, respectively, and $\bar{y}_{R_k}$ is the mean target value within region $k$.

Decision Trees are highly interpretable and capture nonlinear dependencies, but without constraints such as maximum depth or minimum leaf size, they are prone to overfitting.

\subsection{Linear Regression}

Linear Regression assumes a linear relationship between the features and the output. The weight vector $\mathbf{w}$ is computed by minimizing the squared prediction error:

\begin{equation}
\min_{\mathbf{w}} \sum_{i=1}^{n} (y_i - \mathbf{w}^\top \mathbf{x}_i)^2,
\end{equation}

with predictions given by $\hat{y}_i = \mathbf{w}^\top \mathbf{x}_i$. Solutions are obtained analytically or through iterative optimization.

Although computationally efficient and easy to interpret, Linear Regression is limited in its ability to capture nonlinear interactions typical of fading, interference, and modulation dynamics in 5G networks.

\subsection{Light Gradient Boosting Machine (LightGBM)}

LightGBM is a gradient boosting framework optimized for speed and large-scale learning. Unlike level-wise tree expansion, LightGBM grows trees leaf-wise by selecting the split that maximizes gain. The ensemble prediction is given by:

\begin{equation}
\hat{y} = \sum_{b=1}^{B} f_b(\mathbf{x}), \quad f_b \in \mathcal{F}_{\text{leaf-wise}},
\end{equation}

where $f_b$ is the $b$-th boosted tree. LightGBM uses histogram-based binning to reduce memory requirements and accelerate training.

LightGBM is effective for capturing high-dimensional, nonlinear relationships among PHY-layer metrics and provides importance scores to identify influential features.

\subsection{Extreme Gradient Boosting (XGBoost)}

XGBoost constructs tree ensembles sequentially, with each new tree trained to reduce the residual errors of its predecessors. The prediction model takes the form:

\begin{equation}
\hat{y} = \sum_{b=1}^{B} f_b(\mathbf{x}), \quad f_b \in \mathcal{F},
\end{equation}

and uses second-order gradient information (gradient + Hessian) to accelerate optimization. Regularization terms control model complexity and mitigate overfitting.

XGBoost is exceptionally effective for structured 5G dataset modeling, where interactions between coding, scheduling, and SNR metrics create complex feature dependencies.

\subsection{Problem Formulation}

Downlink bit rate and CQI prediction are formulated as supervised regression tasks. The feature sets are defined as:

\begin{equation}
\mathbf{x}_{\text{brate}} = [\text{BLER}, \text{TTI}, \text{MCS}, \text{CQI}],
\end{equation}

\begin{equation}
\mathbf{x}_{\text{CQI}} = [\text{Bit rate}, \text{TTI}, \text{MCS}, \text{BLER}].
\end{equation}

The learning objective is to approximate a function $\hat{y} = f(\mathbf{x})$ that minimizes prediction error. Data is partitioned into 80\% training and 20\% testing. All features undergo z-score normalization using \texttt{StandardScaler}.

Z-score normalization ensures zero mean and unit variance, preventing disproportionate influence from large-scale features and improving optimization stability.

\subsection{Evaluation Metrics}

Model performance is assessed using:

\textbf{(a) Root Mean Squared Error (RMSE)}
\begin{equation}
\text{RMSE} = \sqrt{\frac{1}{n} \sum_{i=1}^{n} (y_i - \hat{y}_i)^2}
\end{equation}

\textbf{(b) Mean Squared Error (MSE)}
\begin{equation}
\text{MSE} = \frac{1}{n} \sum_{i=1}^{n} (y_i - \hat{y}_i)^2
\end{equation}

\textbf{(c) Coefficient of Determination ($R^2$)}
\begin{equation}
R^2 = 1 - 
\frac{\sum_{i=1}^{n} (y_i - \hat{y}_i)^2}{
\sum_{i=1}^{n} (y_i - \bar{y})^2}
\end{equation}

Higher $R^2$ and lower RMSE/MSE indicate better predictive performance.

\subsection{Feature Importance Analysis}

To interpret the decision processes of the models, feature importance values from LightGBM are analyzed. These scores quantify each feature’s contribution to reducing loss across the boosted ensemble and reveal which PHY-layer attributes most strongly influence achievable bit rate in practical 5G deployments.

%% file: sections/system.tex
Our uplink forecasting framework is evaluated on a fully programmable standalone 5G NR deployment designed in alignment with Open RAN principles. The radio access network is realized using the open-source srsRAN 5G protocol stack, providing modular and configurable components for both gNB and core network functionalities. A Dell workstation powered by an Intel i7 CPU and 32 GB RAM operates as the 5G gNB, while core network connectivity is facilitated through a containerized Open5GS deployment. The host machine runs Ubuntu 24.04 LTS with real-time kernel patches to satisfy stringent timing requirements associated with over-the-air transmission in 5G systems.

Radio operations are carried out in the sub–6 GHz band centered at 3.4 GHz using an Ettus Research USRP B210 software-defined radio as the RF front-end. The SDR interfaces to the host through USB 3.0, operating with 20 MHz of bandwidth. Antenna gains and RF parameters are calibrated to ensure reliable connectivity across controlled indoor laboratory trials and outdoor propagation scenarios.

Two Google Pixel 7a smartphones running Android 14 serve as commercial user devices for measurement collection. The UE exposes key physical-layer statistics such as BLER, TTI, bit rate, MCS, and SNR, which form the fundamental features and target variables required for the supervised learning tasks.

\begin{figure}[htbp]
\centering
\includegraphics[width=3.5in,height=1.2in]{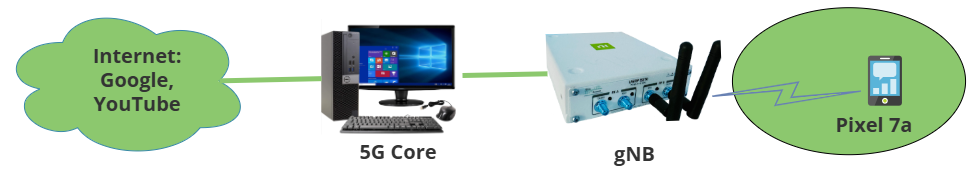}
\caption{Overview of the standalone 5G experimental testbed architecture.}
\label{system-model}
\end{figure}

\subsection{Data Collection Methodology}

To support broad repeatability and scenario diversity, we deployed standalone 5G platforms using both OpenAirInterface and srsRAN across indoor and outdoor environments. All analytical results in this paper are derived from the srsRAN-based platform. As summarized in Table~\ref{data}, experiments were performed under a wide set of operating conditions, including LOS and NLOS environments, static configurations, and mobile user trajectories.

Uplink traffic was generated using a mix of widely used applications—YouTube, Magic iPerf, and Ookla Speedtest—ensuring capture of heterogeneous traffic patterns. This multimodal workload includes sustained flows (iPerf UDP/TCP), burst-based adaptive streaming (YouTube), and commercially deployed throughput estimation (Ookla). Each test was executed multiple times to ensure statistical robustness and to account for channel variability over time.

During all trials, the UE continuously logged PHY-layer statistics, while application-level information was recorded simultaneously. This synchronized measurement approach enables direct correlation between instantaneous radio conditions and achieved throughput. All measurements were timestamped with millisecond precision, allowing fine-grained temporal alignment across layers.

Data was collected across multiple days and environmental settings, producing a comprehensive dataset reflecting a broad span of realistic 5G operational states. These measurements form the basis of our machine learning modeling and validation.

\begin{table}[htbp]
    \centering
    \small
    \caption{Testing scenarios and traffic generations}
    \begin{tabular}{|l|c|c|c|c|}
        \hline
        \textbf{No.} & \textbf{Test Type} & \textbf{Environment} & \textbf{Protocol}& \textbf{Testing Mode}\\
        \hline
        1 & YouTube & Indoor & UDP & Static\\
        \hline
        2 & YouTube & Indoor & UDP & Mobility\\
        \hline
        3 & YouTube & Outdoor & UDP & Static\\
        \hline
        4 & YouTube & Outdoor & UDP & Mobility\\
        \hline
        5 & Ookla & Indoor & TCP & Static\\
        \hline
        6 & Ookla & Indoor & TCP & Mobility\\
        \hline
        7 & Ookla & Outdoor & TCP & Static \\
        \hline
        8 & Ookla & Outdoor & TCP & Mobility\\
        \hline
        9 & iPerf & Indoor & UDP & Static \\
        \hline
        10 & iPerf & Indoor & UDP & Mobility\\
        \hline 
        11 & iPerf & Outdoor & UDP & Static \\
        \hline 
        12 & iPerf & Outdoor & UDP & Mobility\\
        \hline 
        13 & iPerf & Indoor & TCP & Static \\
        \hline
        14 & iPerf & Indoor & TCP & Mobility\\
        \hline 
        15 & iPerf & Outdoor & TCP & Static \\
        \hline 
        16 & iPerf & Outdoor & TCP & Mobility\\
        \hline 
    \end{tabular}
    \label{data}
\end{table}

\subsubsection{Ookla Testing}

Deploying a dedicated Ookla Speedtest server within our laboratory would require allocation of a public IP address so that the server appears in the Speedtest directory. Due to institutional security constraints, public IP assignment requires administrative approval and was not feasible for our experiments. Instead, we relied on both nearby regional servers and distant international endpoints to introduce diverse network paths. We also utilized several iPerf servers located within a 50 km radius to evaluate performance under controlled backhaul conditions. The global Speedtest server pool included nodes in the United Kingdom, Saudi Arabia, Egypt, Ghana, and various North American regions.

\begin{figure}[htbp]
    \centering
    \subfloat[Outdoor Ookla evaluation\label{ookla_outdoor}]{%
        \includegraphics[width=2in,height=2in]{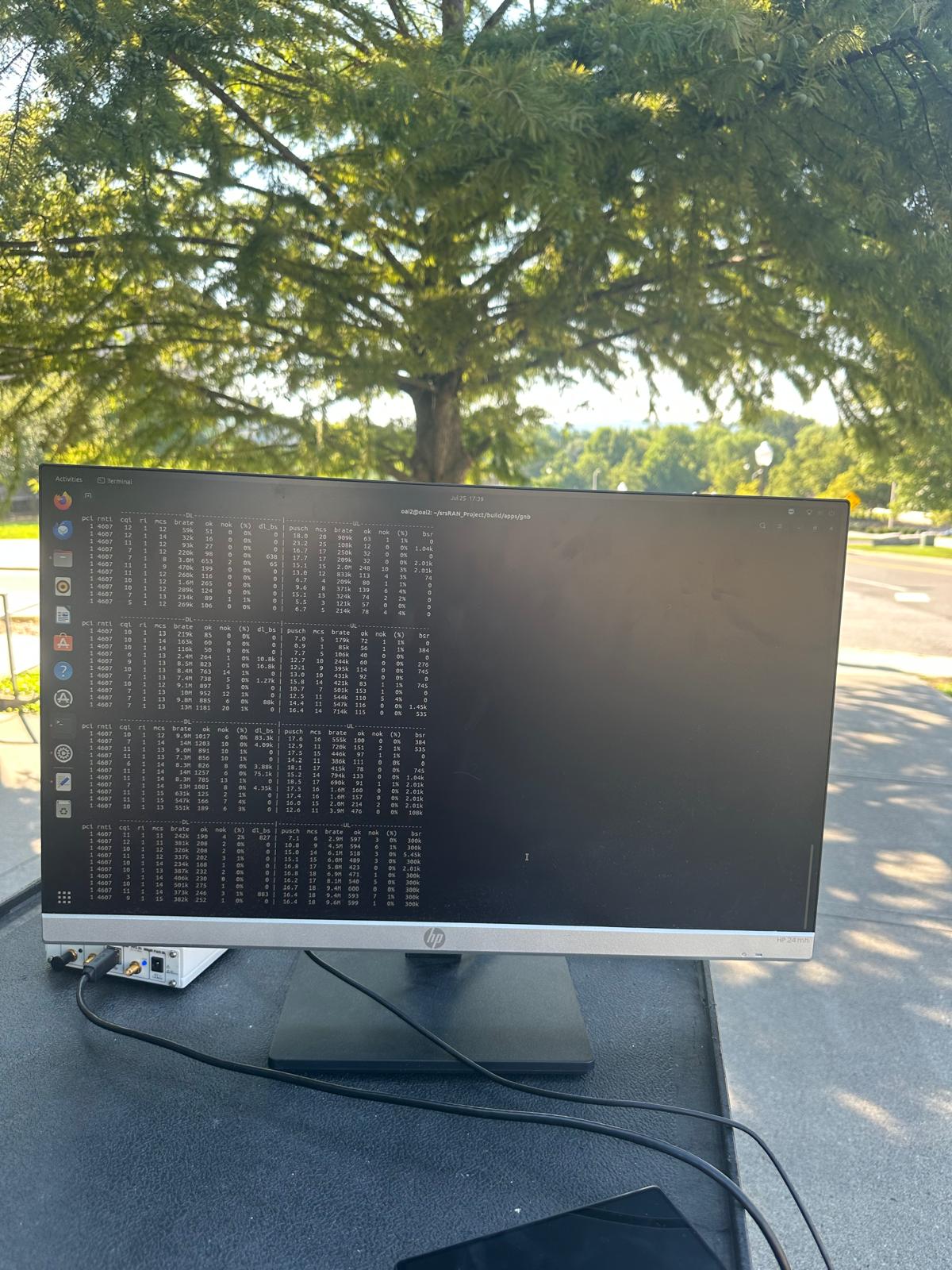}}
    
    \vspace{0.5cm}
    
    \subfloat[Local server (Floyd, VA, USA)\label{ookla_floyd}]{%
        \includegraphics[width=2in,height=2in]{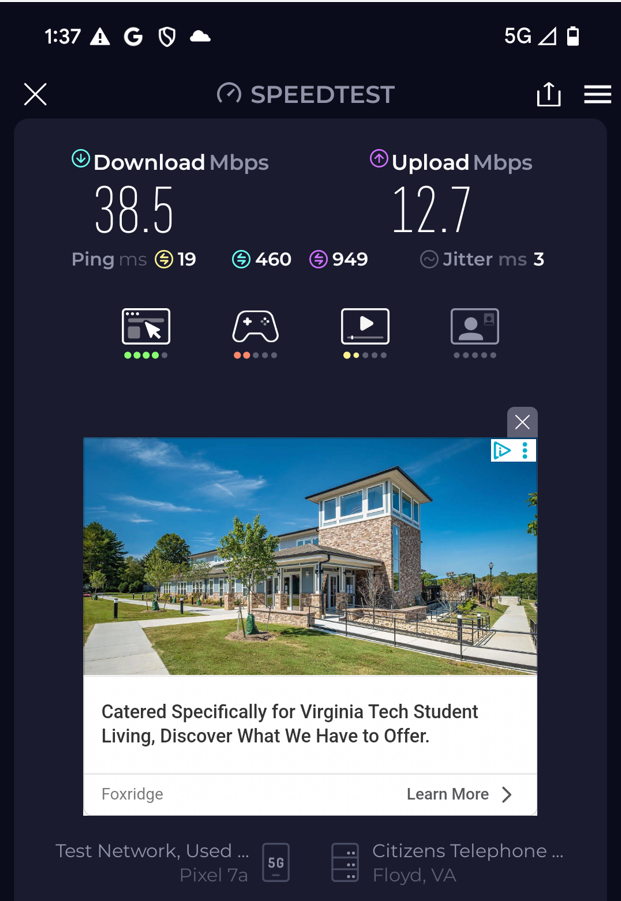}}
    
    \vspace{0.5cm}
    
    \subfloat[International server (Manchester, UK)\label{ookla_manchester}]{%
        \includegraphics[width=2in,height=2in]{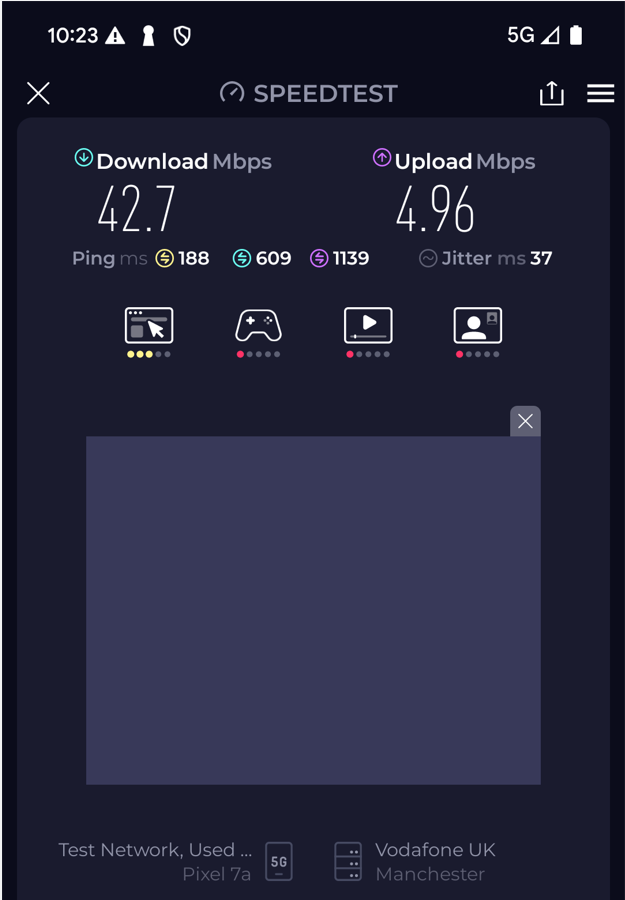}}
    
    \caption{Ookla uplink testing across local and long-distance servers.}
    \label{ookla}
\end{figure}

\subsubsection{iPerf Testing}

Throughput evaluation using the Magic iPerf application on the UE employed both TCP and UDP protocols. To incorporate path diversity, we utilized publicly accessible iPerf servers hosted in Ashburn, VA, and Atlanta, GA. These endpoints allowed us to observe protocol-specific behavior and the effects of backhaul variability on uplink throughput.

\begin{figure}[htbp]
    \centering
    \subfloat[TCP uplink test\label{iperf_tcp}]{%
        \includegraphics[width=2in,height=2in]{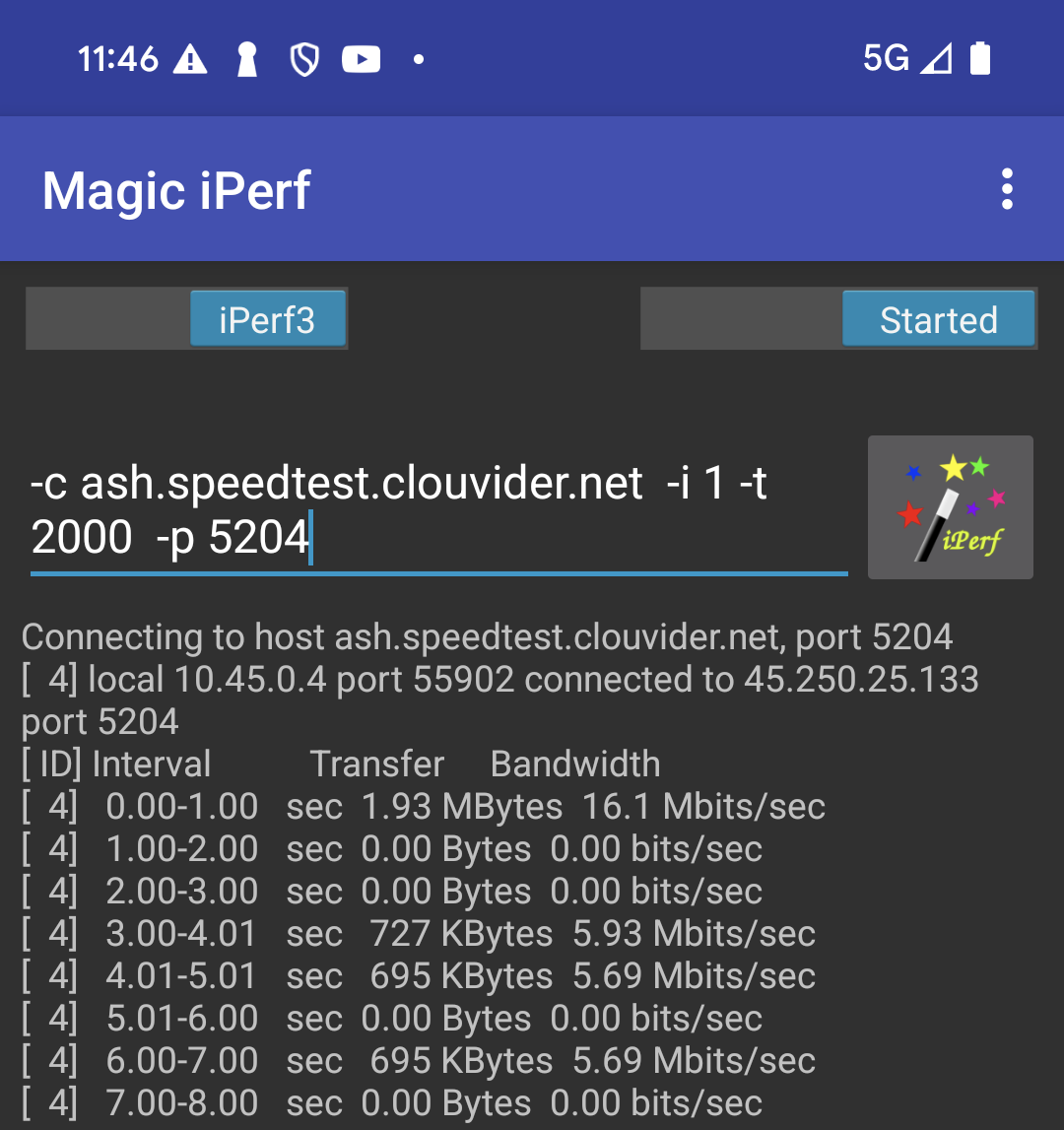}}
    
    \vspace{0.5cm}
    
    \subfloat[UDP uplink test\label{iperf_udp}]{%
        \includegraphics[width=2in,height=2in]{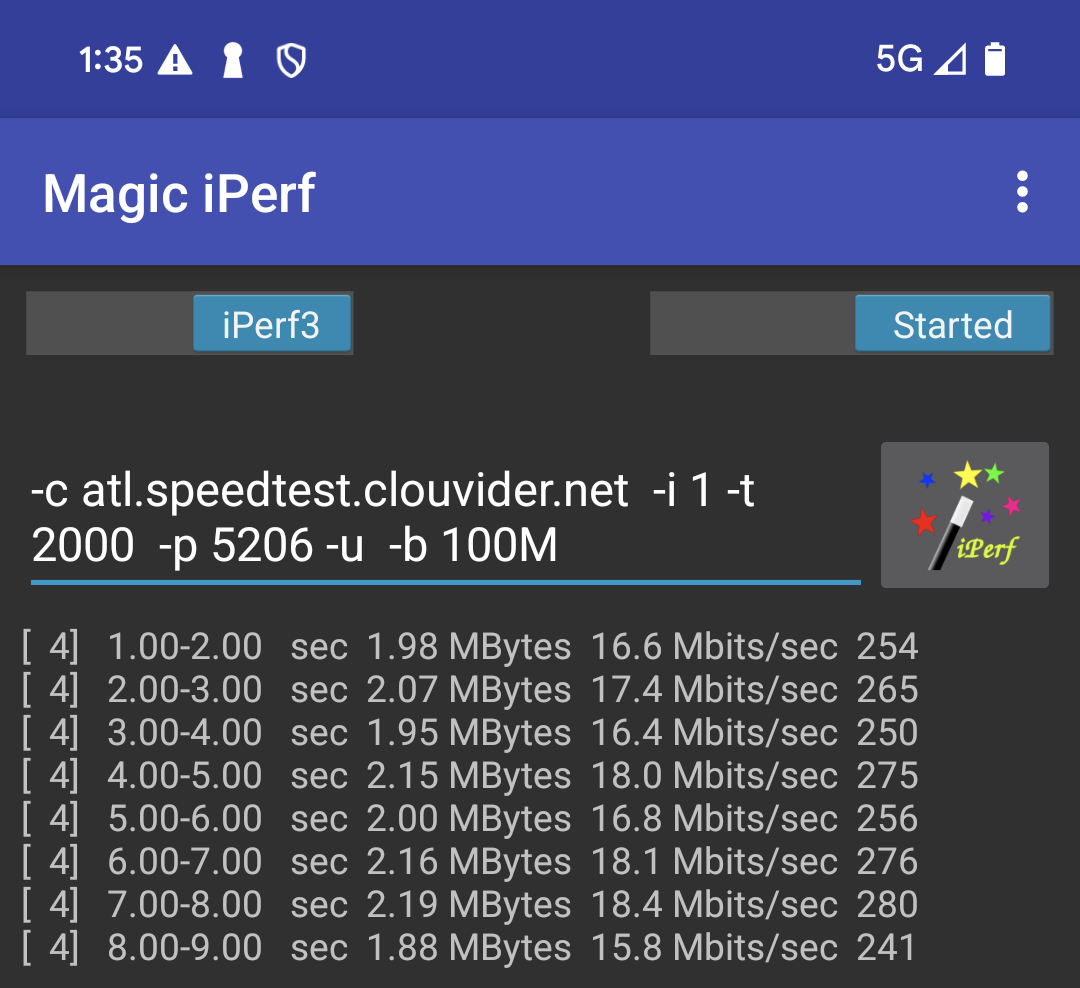}}
    
    \caption{iPerf uplink throughput measurements.}
    \label{iperf}
\end{figure}

\subsubsection{YouTube Testing}

To examine performance under real-world application-driven workloads, we generated traffic using YouTube in both indoor and outdoor configurations. Our experiments focused on uplink-intensive use cases, including 4K and 8K streaming uploads. Unlike the constant-rate behavior of iPerf, YouTube produces highly bursty traffic governed by adaptive buffering and encoding strategies.

\begin{figure}[htbp]
    \centering
    \subfloat[Single UE YouTube testing\label{youtube1}]{%
        \includegraphics[width=2in,height=3in]{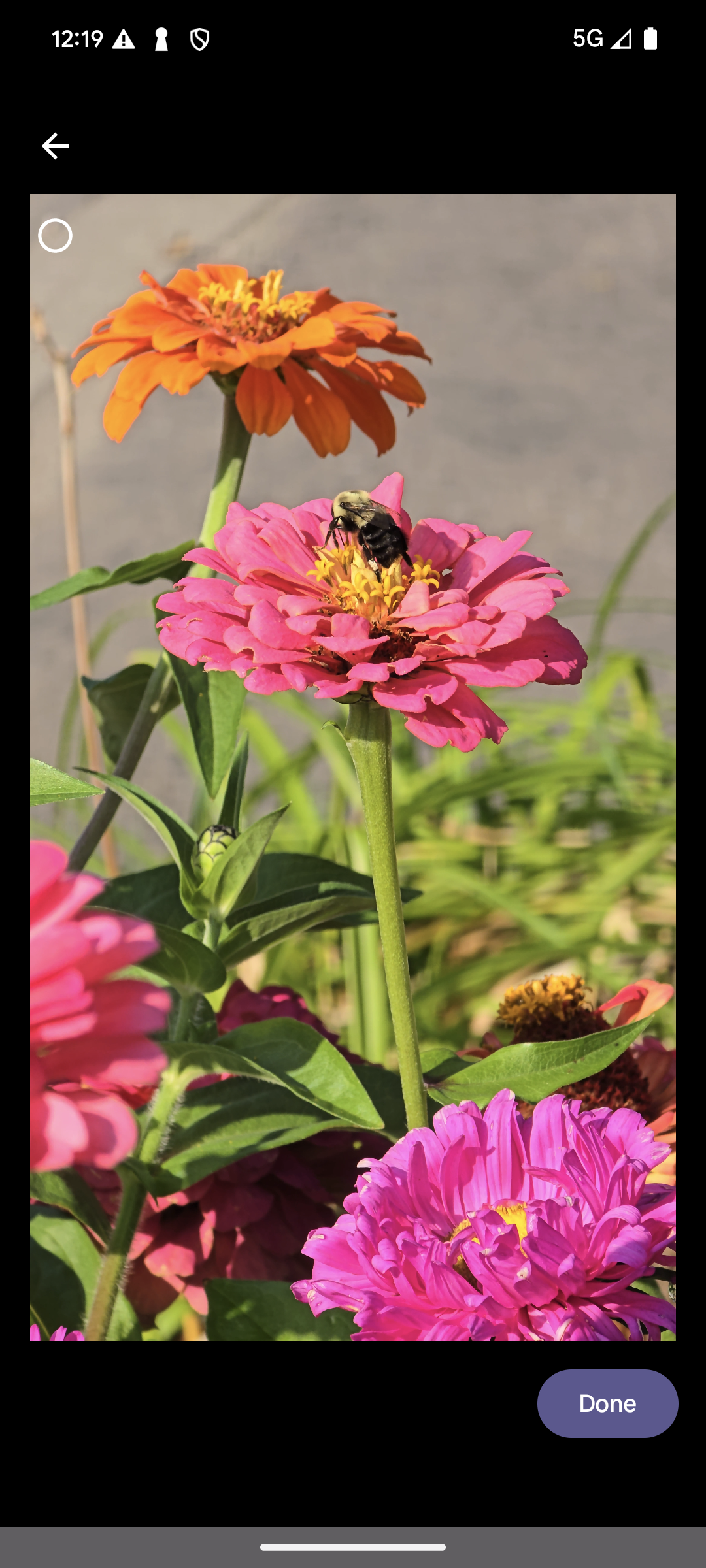}}
    \caption{YouTube-based uplink workload generation.}
    \label{youtube}
\end{figure}

Each record in the dataset constitutes a snapshot of real-time MAC scheduling behavior and the underlying wireless channel state. To prepare the dataset for learning, we applied standard preprocessing steps including timestamp alignment, outlier filtering, and feature normalization.

%% file: sections/simulation.tex
The proposed learning framework was evaluated using five regression models—Linear Regression, Random Forest, Decision Tree, LightGBM, and XGBoost—on the collected 5G SA uplink dataset. Their performance was assessed using three standard regression metrics: Mean Squared Error (MSE), Root Mean Squared Error (RMSE), and the coefficient of determination ($R^2$). The numerical outcomes for bit rate and SNR prediction are summarized in Tables~\ref{table1} and~\ref{table2}, while Fig.~\ref{ml_models} provides a graphical comparison to facilitate visual interpretation of the results.

\subsection{Model Performance Comparison}

\begin{table}[htbp]
    \centering
    \caption{Regression Model Performance for Bit Rate Prediction}
    \begin{tabular}{|l|c|c|c|}
        \hline
        \textbf{Model} & \textbf{MSE ($kbps^2$)} & \textbf{RMSE (kbps)} & \textbf{$R^2$ Score}\\
        \hline
       Linear Regression & 1.292 & 1.137 & 0.951 \\
        \hline
        Decision Tree & 0.138 & 0.371 & 0.995 \\
        \hline
        Random Forest & 0.103 & 0.321 & 0.996\\
        \hline
        XGBoost & 0.099 & 0.314 & 0.996\\
        \hline
        LightGBM & \textbf{0.094} & \textbf{0.306} & \textbf{0.996}\\
        \hline
    \end{tabular}
    \label{table1}
\end{table}

\begin{table}[htbp]
    \centering
    \caption{Regression Model Performance for SNR Prediction}
    \begin{tabular}{|l|c|c|c|}
        \hline
        \textbf{Model} & \textbf{MSE} & \textbf{RMSE} & \textbf{$R^2$ Score}\\
        \hline
        Linear Regression & 6.594 & 2.568 & 0.822 \\
        \hline
        Decision Tree & 5.260 & 2.293 & 0.858 \\
        \hline
        Random Forest & 4.049 & 2.012 & 0.891\\
        \hline
        XGBoost & \textbf{3.758} & \textbf{1.939} & \textbf{0.899}\\
        \hline
        LightGBM & 3.834 & 1.958 & 0.897\\
        \hline
    \end{tabular}
    \label{table2}
\end{table}

\begin{figure}[htbp]
    \centering
    \subfloat[Bit rate performance metrics\label{model_performance1}]{%
        \includegraphics[width=0.8\linewidth,height=5cm]{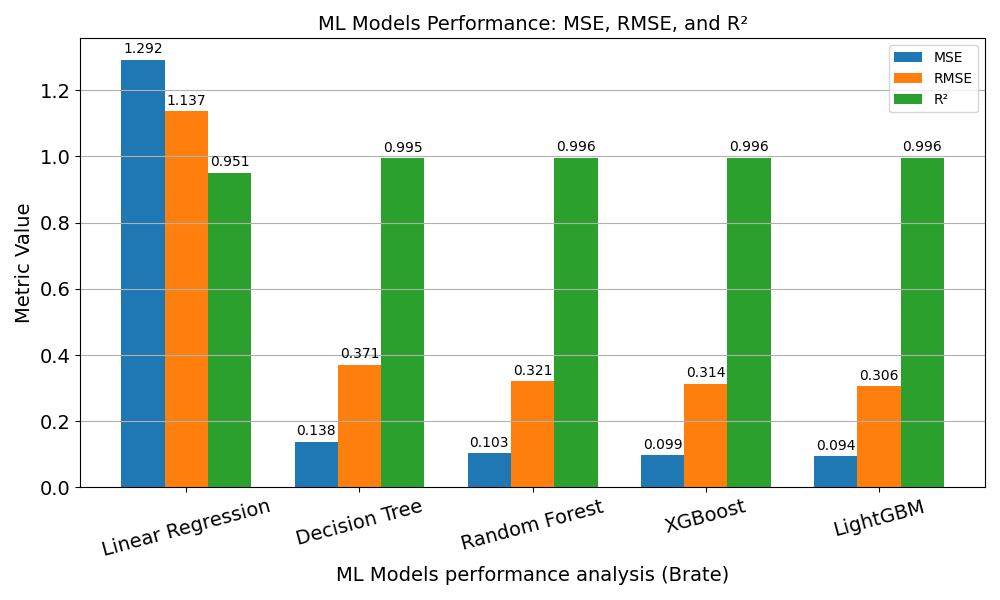}
    }
    
    \vspace{0.5cm}
    
    \subfloat[SNR performance metrics\label{model_performance2}]{%
        \includegraphics[width=0.8\linewidth,height=5cm]{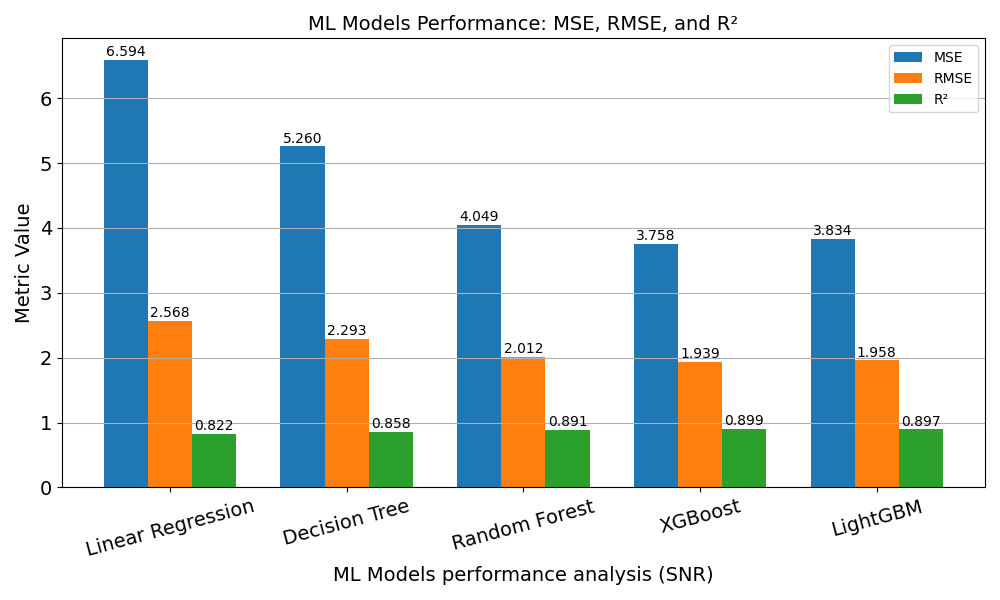}
    }
    \caption{Comparison of regression models across MSE, RMSE, and $R^2$ for both prediction tasks}
    \label{ml_models}
\end{figure}

Figures~\ref{model_performance1} and~\ref{model_performance2} illustrate the comparative behavior of the five regression models along all three evaluation metrics for the bit rate and SNR prediction problems, respectively.

For the bit rate task (Fig.~\ref{model_performance1}), each model attains strong performance with relatively low error and high $R^2$ values. The Linear Regression baseline already achieves $R^2 = 0.951$, indicating that a substantial portion of the variance can be captured through approximately linear relationships among the features. Nonetheless, models capable of learning nonlinear structure provide noticeable gains. Decision Tree Regression improves performance to $R^2 = 0.995$ with MSE = 0.138 $kbps^2$, demonstrating the benefit of capturing nonlinear splits in the feature space. The ensemble-based approaches (Random Forest, XGBoost, and LightGBM) further refine the predictive quality through averaging and boosting mechanisms. LightGBM attains the best overall performance with MSE = 0.094 $kbps^2$, RMSE = 0.306 kbps, and $R^2 = 0.996$, corresponding to roughly a 92.7\% reduction in MSE relative to Linear Regression and a 31.9\% improvement compared to the standalone Decision Tree. The near-unity $R^2$ highlights that LightGBM explains almost all observed variance in bit rate, making it particularly suitable for accurate throughput forecasting.

The SNR prediction problem (Fig.~\ref{model_performance2}) is noticeably more difficult, as reflected in generally lower $R^2$ values across all methods. Linear Regression yields $R^2 = 0.822$, indicating that linear structure alone is insufficient to fully account for SNR variations. Decision Tree and Random Forest models provide improvements, with $R^2 = 0.858$ and $0.891$, respectively, underscoring the value of capturing nonlinear relationships. The gradient boosting models, XGBoost and LightGBM, further enhance accuracy, with XGBoost achieving the best result (MSE = 3.758, RMSE = 1.939, $R^2 = 0.899$). However, the performance gap among the top tree-based methods is narrower for SNR than for bit rate, suggesting that SNR is inherently more challenging to predict from the chosen feature set.

\subsection{Feature Importance Analysis}

\begin{figure}[htbp]
    \centering
    \subfloat[Feature importance for bit rate prediction\label{fig:throughput}]{%
        \includegraphics[width=0.8\linewidth,height=5cm]{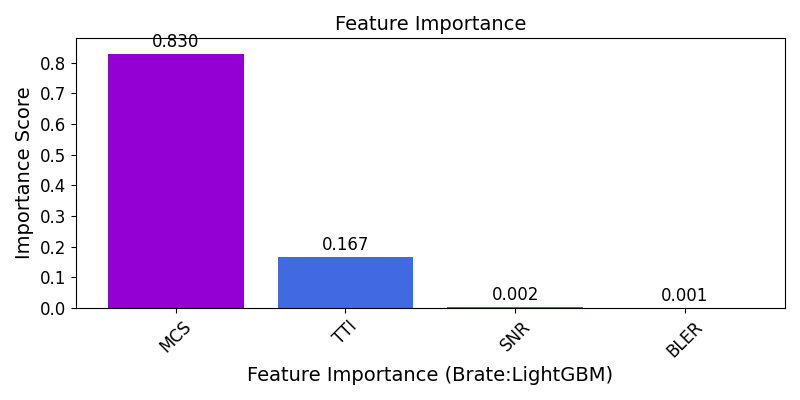}
    }
    
    \vspace{0.5cm}
    
    \subfloat[Feature importance for SNR prediction\label{fig:error}]{%
        \includegraphics[width=0.8\linewidth,height=5cm]{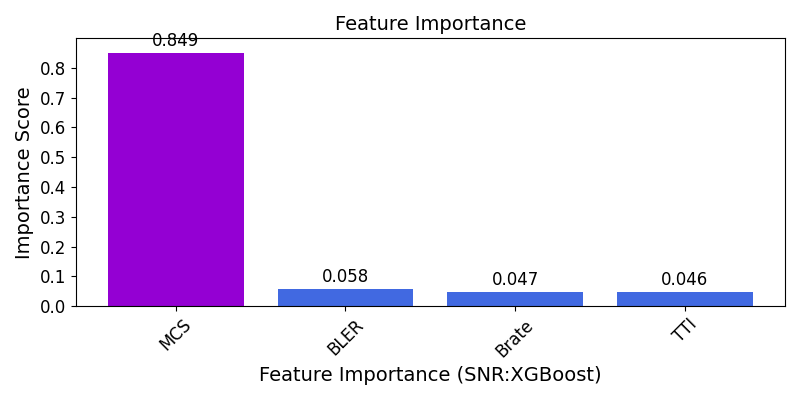}
    }
    \caption{Relative feature importance derived from the LightGBM model}
    \label{features}
\end{figure}

To better understand which physical-layer inputs drive the predictive behavior of the gradient-boosted models, we examine feature importance scores obtained from LightGBM and XGBoost. Figure~\ref{features} displays the relative contributions of each feature.

For bit rate prediction (Fig.~\ref{fig:throughput}), MCS clearly dominates the ranking, contributing the largest share of information gain across the ensemble. This outcome is consistent with the role of MCS in determining spectral efficiency and, consequently, achievable throughput. TTI appears as the second most influential feature, underscoring its impact on the temporal allocation of radio resources and scheduler decisions. In contrast, BLER and SNR provide comparatively limited marginal gain, indicating that, within the given configuration, throughput is chiefly controlled by MAC-layer decisions (MCS and TTI) rather than raw channel quality indicators alone. These insights suggest that ML-assisted scheduling policies should emphasize accurate modeling and adaptation of MCS and TTI settings to maximize bit rate performance.

The feature importance pattern for SNR prediction (Fig.~\ref{fig:error}) differs noticeably. Again, MCS occupies the top position, with higher contribution than the remaining features. This result indicates a strong coupling between selected MCS levels and the underlying SNR, reflecting the scheduler’s tendency to adapt MCS in response to channel conditions. The remaining three inputs—bit rate, TTI, and BLER—exhibit similar importance scores, together providing complementary information about link quality dynamics. Overall, the analysis reveals that scheduling and modulation decisions are tightly intertwined with the effective SNR observed at the UE.

\subsection{Prediction Accuracy Visualization}

\begin{figure}[htbp]
    \centering
    \subfloat[Actual vs. predicted bit rate\label{actual_vs_predicted1}]{%
        \includegraphics[width=0.8\linewidth,height=5cm]{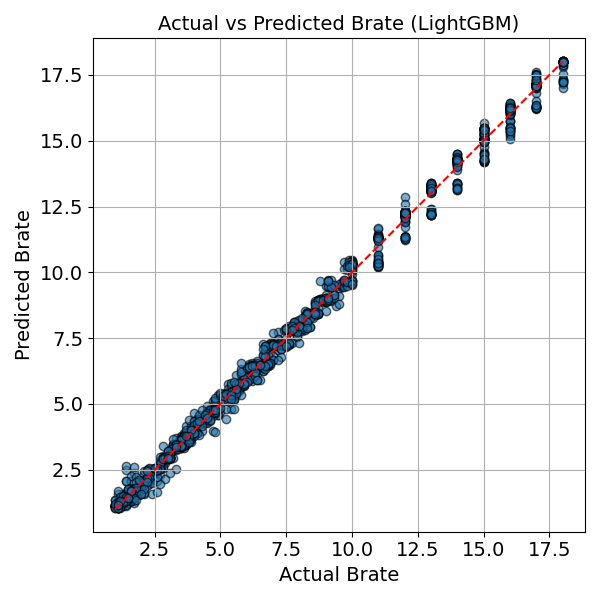}
    }
    
    \vspace{0.5cm}
    
    \subfloat[Actual vs. predicted SNR\label{actual_vs_predicted2}]{%
        \includegraphics[width=0.8\linewidth,height=5cm]{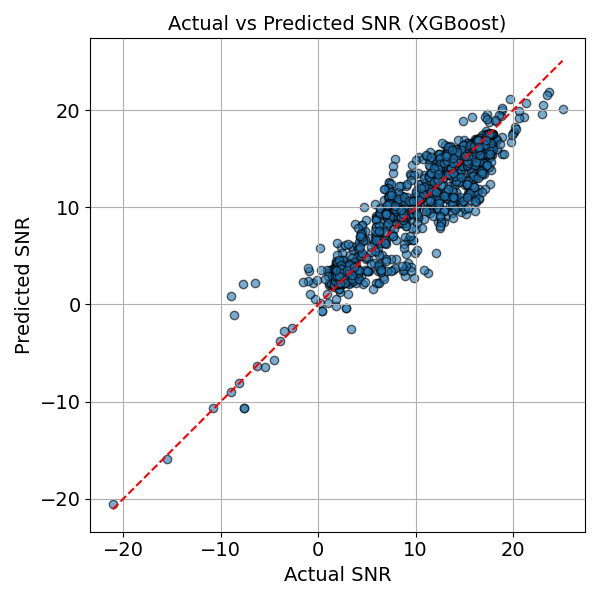}
    }
    \caption{Scatter plots comparing LightGBM predictions with ground truth}
    \label{actual_vs_predicted}
\end{figure}

Figures~\ref{actual_vs_predicted1} and~\ref{actual_vs_predicted2} visualize the relationship between predicted and measured values for LightGBM (and XGBoost, if overlaid), using scatter plots. The red dashed diagonal in each plot corresponds to the ideal case where $\hat{y} = y$.

For the bit rate task (Fig.~\ref{actual_vs_predicted1}), the vast majority of points lie tightly clustered along the diagonal across the full range of approximately 1--10~Mbps. This indicates that the model maintains high fidelity in both low and high throughput regimes, with minimal dispersion around the ideal line. The symmetry of deviations above and below the diagonal suggests that the model is well-calibrated, exhibiting no systematic overestimation or underestimation trends.

For SNR prediction (Fig.~\ref{actual_vs_predicted2}), samples also align reasonably well with the diagonal over the typical operating range (e.g., 0--20~dB), but with visibly larger spread compared to the bit rate case. A slight tendency toward underestimation at lower SNR values can be observed, hinting that some residual nonlinear effects or unmodeled factors remain. Nevertheless, the overall alignment confirms that the gradient-boosted models capture the dominant structure of SNR variation from the available features.

\subsection{Prediction Error Distribution Analysis}

\begin{figure}[htbp]
    \centering
    \subfloat[Bit rate prediction error histogram\label{prediction_error_distribution1}]{%
        \includegraphics[width=0.8\linewidth,height=5cm]{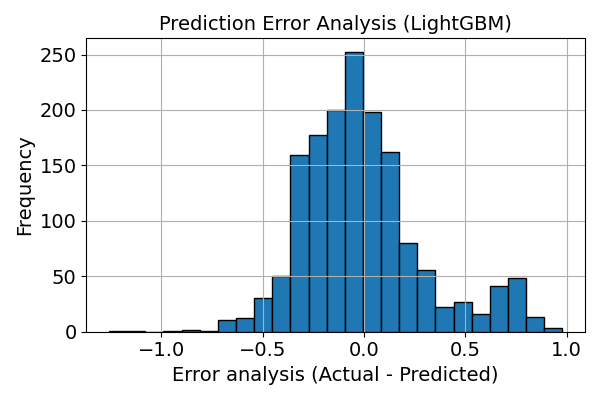}
    }
    
    \vspace{0.5cm}
    
    \subfloat[SNR prediction error histogram\label{prediction_error_distribution2}]{%
        \includegraphics[width=0.8\linewidth,height=5cm]{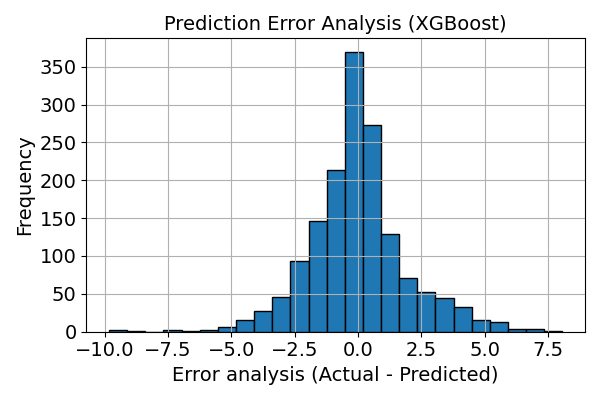}
    }
    \caption{Histogram of prediction errors for LightGBM models}
    \label{prediction_error_distribution}
\end{figure}

Figures~\ref{prediction_error_distribution1} and~\ref{prediction_error_distribution2} show the empirical distributions of prediction errors, defined as $\varepsilon = y_{\text{actual}} - y_{\text{predicted}}$, for bit rate and SNR, respectively.

The bit rate error histogram (Fig.~\ref{prediction_error_distribution1}) is sharply concentrated around zero with a shape close to Gaussian, and most errors fall within $\pm 2.5$~Mbps. The tails are relatively light, and the distribution exhibits little skew, indicating that positive and negative errors are roughly balanced. Only a small fraction of samples deviate by more than $\pm 5$~Mbps, suggesting strong robustness even in challenging cases. This behavior implies that the residual error is primarily driven by intrinsic randomness in the wireless channel and measurement noise rather than systematic model bias. Such properties are beneficial for downstream tasks that rely on confidence estimation or uncertainty bounds.

In contrast, SNR prediction errors (Fig.~\ref{prediction_error_distribution2}) span a broader range, with most values lying within approximately $\pm 5$ units. While this appears larger in absolute terms, it must be interpreted in light of SNR’s narrower dynamic range. When normalized by the actual SNR span, the relative prediction quality is comparable to that of the bit rate model. The wider spread reflects the inherent difficulty of inferring SNR solely from higher-layer performance indicators and scheduling parameters, yet the error distribution remains centered close to zero, confirming the absence of systematic bias.

%% file: sections/conclusion.tex
We propose a learning-based prediction architecture for estimating uplink SNR and bit rate in 5G NR networks by leveraging real-world measurements gathered from two smartphones operating on an srsRAN platform. The dataset comprises PHY-layer indicators—SNR, MCS, TTI, BLER, and bit rate—collected over diverse propagation conditions such as LOS, nLOS, indoor, outdoor, static, and mobile scenarios. Using MSE, RMSE, and $R^2$ as evaluation metrics, five regression methods were compared. LGBM emerged as the most effective model for bit-rate prediction, whereas XGBoost delivered superior SNR estimation, reflecting the suitability of boosted trees for capturing nonlinear wireless effects. Feature-importance analysis identified MCS and TTI as the most critical inputs. The resulting framework enables practical QoS prediction using only radio-side measurements and supports dynamic resource-management strategies. The joint SNR–bit-rate prediction capability also provides a principled basis for constructing RL state representations to be used in our subsequent research.